\documentclass[twocolumn,prb,english,showpacs,preprintnumbers,amsmath,amssymb]{revtex4}
\usepackage[latin1]{inputenc}
\usepackage{amsmath}
\usepackage{babel}
\usepackage{graphics}
\usepackage{amssymb}

\newcommand{\up}{\uparrow}
\newcommand{\dn}{\downarrow}

\begin{document}


\title{Modulation of charge-density waves by superlattice structures}

\author{Andr\'e L.\ \surname{Malvezzi},$^1$ 
        Thereza \surname{Paiva},$^2$
        and 
        Raimundo R. \surname{dos Santos}$^2$} 

\affiliation{$^{1}$Departamento\ de F\'\i sica,
                 Faculdade de Ci\^encias,
                 Universidade Estadual Paulista,
                 Caixa Postal 473,
                 17015-970 Bauru SP,
                 Brazil\\
             $^2$Instituto de F\'\i sica, 
                 Universidade Federal do Rio de Janeiro, 
                 Caixa Postal 68528,
                 21941-972 Rio de Janeiro RJ, 
                 Brazil}

\begin{abstract}
We discuss the interplay between electronic correlations and an underlying superlattice structure in determining the period of charge density waves (CDW's), by considering a one-dimensional Hubbard model with a repeated (non-random) pattern of repulsive ($U>0$) and free ($U=0$) sites. Density matrix renormalization group diagonalization of finite systems (up to 120 sites) is used to calculate the charge-density correlation function and structure factor in the ground state. The modulation period can still be predicted through effective Fermi wavevectors, $k{_F}^*$, and densities, and we have found that it is much more sensitive to electron (or hole) doping, both because of the narrow range of densities needed to go from $q^*=0$ to $\pi$, but also due to sharp $2k{_F}^*$-$\,4k{_F}^*$ transitions; these features render CDW's more versatile for actual applications in heterostructures than in homogeneous systems.
\end{abstract}

\date{\today}


\maketitle

The appearance of charge-density waves (CDW's) in conductors with a chain-like structure is a well known phenomenon:\cite{Pouget89,Gruner94,Brown94} electrons condense into a ground state with a modulated charge distribution, as opposed to the uniform distribution of normal metals. 
Depending on their modulation period, these CDW's may slide along the chains to exhibit a wealth of interesting transport phenomena such as non-ohmic behavior, and AC current oscillations  induced by DC electric fields.\cite{Pouget89,Gruner94,Brown94} Nonetheless, the fact that CDW transport was only controlled in bulk samples has hindered its integration to devices for some time. However, advances in material synthesis and manipulation over the last decade or so have given rise to new, micro- and nanostrucured, CDW systems. For instance, films of Rb$_{0.30}$MoO$_3$ (``blue bronze'') with thickness in the range 100-1000 nm have been grown by pulsed-laser deposition;\cite{vanderZant96} mesoscopic wire structures have also been obtained.\cite{Mantel99} Also, regions of negative absolute resistance have been observed at sufficiently low temperatures in single crystals of NbSe$_3$ and $o$-TaS$_3$ with cross sections of 0.2 to 1 $\mu$m$^2$.\cite{Mantel00,vanderZant01} These approaches open up the possibility of fabricating CDW heterostructures, the properties of which are still unknown, and certainly worth investigating.

While most of the above-mentioned features of CDW systems in confined geometries are dynamical, the knowledge of static properties is also of primary importance. From the theoretical point of view, the modulation period of the CDW is one of the key issues, even in the case of homogeneous systems. This has been addressed with the aid of various models, the simplest one being the one-dimensional Hubbard model, but the modulation period is still a matter of debate. On the one hand, the Luttinger liquid (LL) description\cite{Solyom79,Schulz90,Frahm90,Voit94} predicts that charge correlations should be dominated by a wavevector $q^*=2k_F$, where $k_F=\pi\rho/2$ [$k_F=\pi(2-\rho)/2$] is the Fermi wave vector for an electronic density $\rho<1$ [$\rho>1$] of free electrons on a periodic lattice. By contrast, numerical analyses predict that the CDW instability is dominated by the $4k_F$ term, for strong enough on-site repulsion, $U$.\cite{Hirsch83a,Hirsch83b,Hirsch84,Paiva00a}
These two scenarios can be reconciled if the amplitude of the $2k_F$ contribution vanishes above some crossover coupling, $U^\times(\rho)$.
At any rate, considerable insight can be gained by examining the behavior of CDW's in a special class of heterostructures. Indeed, a simple model to study the interplay between electronic correlations and an underlying superlattice structure has been proposed;\cite{Paiva96} its Hamiltonian reads
\begin{equation} 
\label{Ham} 
{\cal H}=-t\sum_{i,\sigma}
\left(c_{i\sigma}^{^{\dagger}} c_{i+1\sigma}+\text{H.c.}\right)  + \sum_i
U_i\ n_{i\uparrow}n_{i\downarrow} 
\end{equation} 
where, in standard notation, $i$ runs over the sites of a one-dimensional
lattice, $c_{i\sigma}^{^\dagger}$ ($c_{i\sigma}$) creates (annihilates) a
fermion at site $i$ in the spin state $\sigma=\uparrow\ {\rm or}\
\downarrow$, and 
$n_{i\sigma}=c_{i\sigma}^{^\dagger}c_{i\sigma}$. 
The on-site Coulomb repulsion is taken to be site-dependent, in a repeated pattern of $L_U$ sites with $U_i=U>0$ (the repulsive layer), followed by $L_0$ sites with $U_i=0$ (the free layer); one defines the `aspect ratio' 
as $\ell=L_U/L_0$. 

The magnetic properties of the above model turned out to be quite
different from those of the corresponding homogeneous
system.\cite{Paiva96,Paiva00b,Malvezzi02} Also, the density at which the system is a Mott insulator, $\rho_I$, is shifted from half filling to a value which depends on the aspect ratio.\cite{Paiva98} These findings on a discrete lattice have been confirmed by an  
extension of the above model to a superlattice made up of Luttinger liquids (LL).\cite{Silva-Valencia01,Silva-Valencia02}

As far as charge-density waves in these Hubbard superlattices are concerned, a preliminary account of their behavior has been presented in Ref.\ \onlinecite{Paiva02}; by virtue of the limitations imposed by Lanczos diagonalizations, that study was restricted to the electronic density $\rho=11/6$ on lattices with $N_s\leq 24$ sites. 
Nonetheless, the results were quite surprising: it was found that the CDW wavevector can be cast in a simple form, $q^*=4k{_F}^*$, where the effective Fermi wavevector, $k{_F}^*$, depends on the density $\rho$, on $L_U$, and on $L_0$ (see below); also, this CDW wavevector oscillates with the number of spacer (i.e., free) sites, with a period which is half of that for the \emph{spin}-density--wave for the same case. In view of this, many interesting questions arise, such as how does $q^*$ depend on the electronic density and on the Coulomb repulsion $U$, and if there are sharp $2k{_F}^*$-$4k{_F}^*$ transitions. In order to answer these questions one needs to examine both a wide range of band fillings and large lattices, so here we report on a Density Matrix Renormalization Group (DMRG) approach\cite{White92,White93,Malvezzi03,Dukelsky04} to this problem. 

\begin{figure}[t]
{\centering\resizebox*{8.8cm}{!}{\includegraphics*{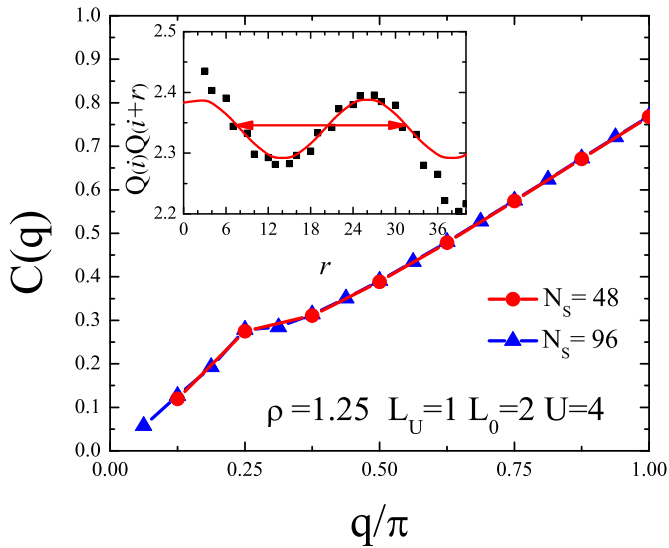}}}
\caption{(Color online) The charge-density structure factor, for two different system sizes, $N_s=48$ (circles) and 96 (triangles), and for the same density: the cusp is located at $q^*=\pi/4$, irrespective of $N_s$. The inset shows its Fourier transform, the correlation function, for the 48-site superlattice, with $r=0$ taken at a free site: the curve guides the eye through the oscillatory part, whose period is marked by the double arrow.}
\label{fig1}
\end{figure}

We consider the Hamiltonian (\ref{Ham}) on lattices with $N_s$ sites, $N_e$
electrons, and open boundary conditions.  The appropriate Finite
Size Scaling (FSS) parameter, however, is the number of periodic cells,
$N_c=N_s/N_b$, for a lattice basis with $N_b=L_U + L_0$ sites. The ground-state, 
$| \psi_0 \rangle $, and energy are obtained through the DMRG method. 
Lattice sizes ranged from 48 to 120 sites and truncation errors in the DMRG 
procedure were kept around $10^{-5}$ or smaller. We have performed a systematic 
study of the charge-density structure factor, 
\begin{equation}
C(q)=\frac{1}{N_c} \sum_{i,j} {\rm e}^{iq(r_i-r_j)}
\langle  Q_i Q_j \rangle\;,
\label{Cq}
\end{equation}
where
$\langle  Q_i Q_j \rangle= \langle \psi_0|  n_i n_j
|\psi_0\rangle$
with $n_i=n_{i\uparrow}+n_{i\downarrow}$;  
the signature of a CDW instability is a cusp in $C(q)$ at $q=q^*$. 
(We chose to examine the charge-density structure factor without the uncorrelated part removed; in so doing, $C(0)=N_e^2/N_c$, instead of satisfying the sum rule $C(0)=0$, but we benefit from the fact that the cusps are much sharper, and thus more readily located.)
We have considered different values of the Coulomb repulsion $U$, 
different occupations $\rho = N_e/N_s$ and different configurations $\{U_i\}$. Not all
configurations $\{U_i\}$ fit into all sizes and occupations but, since
DMRG allows us to study a wide range of lattice sizes, we were able to
establish overall trends.

A typical plot of $C(q)$ is shown in Fig.\ \ref{fig1} for lattices with 48 and 96 sites, $L_U=1$, $L_0=2$, and for density 5/4: a cusp is observed at $q^*=\pi/4$. It is also evident from the figure that neither the cusp position nor the cusp height depend on the system size; thus, our analyses are not plagued by subtle finite-size effects.
The inset of Fig.\ \ref{fig1} displays the charge-density correlation function in the case of $N_s=48$; only the correlation between charges on free sites are plotted, in order to separate intercell from intracell oscillations. We see that once boundary effects die out (typically beyond three unit cells) an oscillatory regime sets in with period $\lambda^*\equiv 2\pi/q^*=8$ unit cells.   

\begin{figure}[t]
{\centering\resizebox*{8.8cm}{!}{\includegraphics*{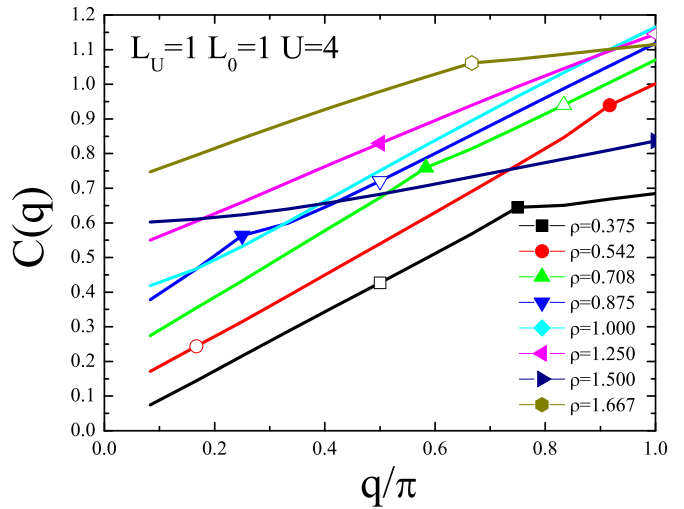}}}
\caption{(Color online) The charge-density structure factor at different electron densities, for a superlattice with 24 cells (48 sites): in each case, filled and empty symbols respectively mark $2k{_F}^*$ and $4k{_F}^*$ (see text), and each plot has been shifted vertically by the same amount of 0.1, for clarity.}
\label{fig2}
\end{figure}

Before comparing the behavior at different electron densities, 
we recall some of its special values which play a crucial role in establishing different 
regimes.\cite{Malvezzi02,Paiva02} First, $\rho_0\equiv 1/(L_U+L_0)$ signals the presence of one electron per cell. For $U$ large enough, added electrons will accumulate on the free sites; this saturates at $\rho_{\up\dn}\equiv 2/(1+\ell)$. As the density increases further, one eventually reaches the Mott-Hubbard regime\cite{Paiva98,Malvezzi02,Paiva02} at $\rho_I \equiv {(2+\ell)/(1+\ell)}$, corresponding to the addition of one electron to every repulsive site. Also, the \emph{cell} electronic density is 
\begin{equation}
\rho_{\rm cell}=
\begin{cases}
N_e/N_c       & \text{if\ } \rho < \rho_{\up\dn}\\
N_e/N_c-2L_0  & \text{if\ } \rho > \rho_{\up\dn}.
\end{cases}
\label{rhocell}
\end{equation}

Figure \ref{fig2} shows $C(q)$ for the $L_U=L_0=1$ superlattice for a wide range of densities, all with $U=4$. The cusps move as the density varies, but two regimes can be clearly distinguished: for $\rho < \rho_{\up\dn}=1$, the cusps occur at $q^*=2k{_F}^*$, whereas for $\rho > \rho_{\up\dn}=1$, they are at $q^*=4k{_F}^*$, where the effective Fermi wave vector, $k_F^*$, is defined as 
\begin{equation}
2k_F^*=
\begin{cases}
\pi \rho_{\rm cell} & \text{if\ } \rho_{\rm cell} < 1\\
\pi (2- \rho_{\rm cell}) & \text{if\ } \rho_{\rm cell} > 1.
\label{kfeff}
\end{cases}
\end{equation}

\begin{figure}[t]
{\centering\resizebox*{8.6cm}{!}{\includegraphics*{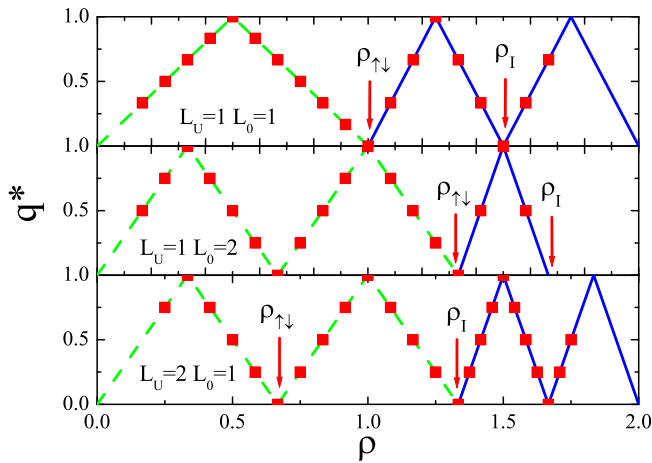}}}
\caption{(Color online) The cusp position as a function of the electrinc density, for three representative superlattice configurations, all with $U=4$: (a) for $L_U=L_0=1$, (b) for $L_U=1,\ L_0=2$, and for $L_U=2,\ L_0=1$. The points have been determined through analyses of $C(q)$, while the dashed and full straight lines respectively represent $2k_F^*$ and $4k_F^*$ [see Eq.\ (\ref{kfeff})].}
\label{fig3}
\end{figure}

The above analysis was repeated for different superlattice configurations, and we followed the evolution of the cusp positions with the electron density. Some of the results (still for $U=4$) are summarized in the $q^*$ \emph{vs.} $\rho$ plots shown in Fig.\ \ref{fig3}. The signature of the $2k{_F}^*$-$\,4k{_F}^*$ transition is now a doubling of the slope.
For this value of $U$ and for $L_U\leq L_0$, the transition occurs at $\rho = \rho_{\up\dn}$, which can be understood by noticing that electrons occupy the free sites for densities below $\rho_{\up\dn}$; above this density, additional electrons occupy the repulsive sites, thus forcing an overall charge rearrangement in order to minimize the on-site repulsion.  
For $L_U > L_0$, on the other hand, $\rho_{\up\dn}<1$ so that electrons can occupy repulsive sites at no cost in energy until one reaches $\rho = \rho_I$; in this case, it is only for  densities larger than $\rho_I$ that a major charge distribution becomes imperative. Another important feature of a superlattice structure is that all modulations become accessible within a range of dopings smaller than in homogeneous systems, for which $q^*=\pi\rho$ or $2\pi\rho$, corresponding to $2k_F$ or $4k_F$, respectively.

\begin{figure}[t]
{\centering\resizebox*{8.4cm}{!}{\includegraphics*{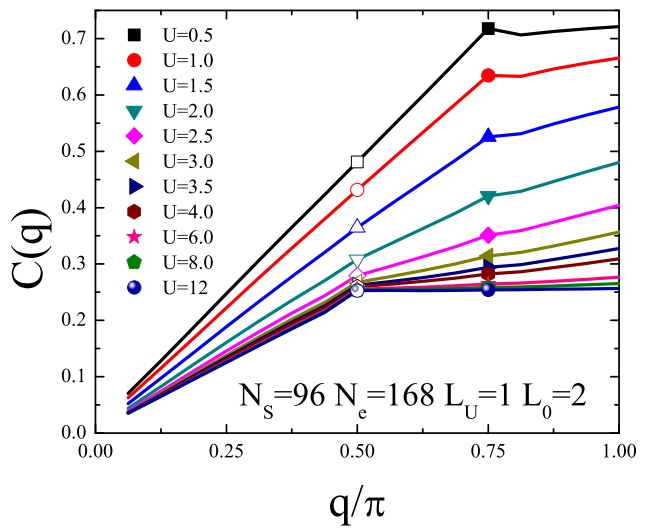}}}
\caption{(Color online) The charge-density structure factor at a fixed electron density ($\rho =7/4$), and for several values of $U$, for a superlattice with 32 cells (96 sites): in each case, filled and empty symbols respectively mark $2k{_F}^*$ and $4k{_F}^*$.}
\label{fig4}
\end{figure}

While the scenario of Fig.\ \ref{fig3} is generic for all $U>4$, the same does not hold for $U\alt 3$. Figure \ref{fig4} shows $C(q)$ for several values of $U$, at a fixed density $\rho > \rho_{\up\dn}$: we see that the cusps move from $2k{_F}^*$ to $4k{_F}^*$ as $U$ is increased. As the density is further increased, the $2k{_F}^*$-$\,4k{_F}^*$ transition occurs at smaller values of $U$; indeed, at weak coupling the occupation of repulsive sites can be compensated by delocalizing the electrons. 

\begin{figure}[h]
{\centering\resizebox*{8.0cm}{!}{\includegraphics*{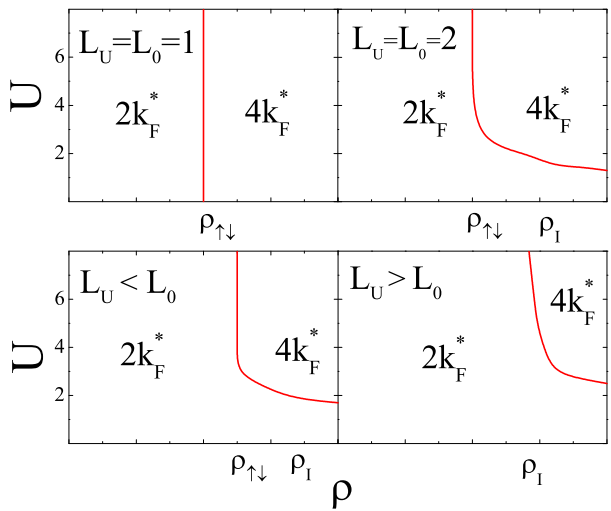}}}
\caption{(Color online) Schematic phase diagram for the $2k{_F}^*$-$\,4k{_F}^*$ transition in the space $U\ vs.\ \rho$: The top two panels correspond to superlattices with the same aspect ratio, $\ell = 1$, while the bottom two are for $\ell < 1$ (left) and $\ell > 1$ (right).}
\label{fig5}
\end{figure}

The picture that emerges is then summarized in Fig.\ \ref{fig5}: with the exception for the case $L_U=L_0=1$, for all other superlattices the transition continuously migrates to larger densities, along a phase boundary line, $U_c(\rho)$. 
It should be noted that the transition boundary for $\rho < \rho_I$, in the case of $L_U>L_0$, is not as steep as that for $L_U < L_0$ and $\rho<\rho_{\up\dn}$; for intance, for the specific case of $L_U=2$ and $L_0=1$, we estimate $U_c\sim 20$, 8, and 3, for $\rho=7/6$, 5/4, and 17/12, respectively.

For consistency, we have also examined the Luttinger liquid (LL)\cite{Voit94,Miranda03} critical exponent, $K_\rho$, which is related\cite{Daul98} to $C(q)$ through
\begin{equation}
K_\rho=\pi \frac{\partial C(q)}{\partial q}\biggr|_{q=0}.
\label{Krho}
\end{equation}
From a plot similar to that of Fig.\ \ref{fig2} (i.e., one for the same parameters, but with the uncorrelated part removed), we see that the slope at $q=0$ first increases with $\rho$, reaches a maximum value and then decreases to zero at $\rho_I$; the actual values of $K_\rho$ are in excellent agreement with those obtained in the LL approach to this problem.\cite{Silva-Valencia01,Silva-Valencia02} By the same token, we have found that $K_\rho$ steadily decreases with $U$ at fixed density, similarly to the homogeneous LL.

In summary, the present study 
has established that superlattices made up of normal metals and of charge-density--wave materials lead to a wider choice of outcomes for the CDW period than in homogeneous systems: the modulation period is much more sensitive to electron (or hole) doping, both because of the narrow range of densities needed to go from $q^*=0$ to $\pi$, but also due to the sharp $2k{_F}^*$-$\,4k{_F}^*$ transitions. The different nature of the sites within each lattice cell (i.e., repulsive of free) leaves room for intracell charge rearrangements, in addition to the intercell ones, the latter being at closer correspondence with the homogeneous systems. Indeed, between intermediate and strong couplings, these transitions are density driven, and occur when the electronic density is increased beyond one electron per free site, if there are more free sites than repulsive ones; if there are more repulsive sites, the transitions take place near the Mott-Hubbard insulating density. For weak couplings, and if the number of repulsive and free sites are different, the superlattice structure plays a secondary role in determining the modulation: the critical on-site repulsion displays just a weak dependence with the density; 
in this regime, the $2k{_F}^*$-$\,4k{_F}^*$ transition may be driven by applying pressure, which increases $U$. 
Therefore, from the standpoint of applications, CDW's seem to be more versatile in heterostructures than in homogeneous systems, and experimental work aiming at assembling CDW materials along these lines should unveil a variety of new effects and properties.

\acknowledgments 
Financial support from the Brazilian Agencies FAPESP, FAPERJ, CNPq (through the Millenium Institute for Nanosciences and the Brazilian Network for Nanostructured Materials initiatives, as well as through fellowships and grants), and FUJB is gratefully acknowledged.

\bibliography{biblio-cdwsl}

\end{document}